\documentclass[preprint,showpacs,preprintnumbers,amssymb,nofootinbib]{revtex4}

\usepackage{graphicx}
\usepackage{bm}

\begin{document}

\centerline{}
\vskip 4cm
\title{Gravitational Radiation in $\bm D$-dimensional Spacetimes}

\author{Vitor Cardoso}
\email{vcardoso@fisica.ist.utl.pt}
\author{\'Oscar J. C. Dias}
\email{oscar@fisica.ist.utl.pt}
\author{Jos\'e P. S. Lemos}
\email{lemos@kelvin.ist.utl.pt}
\affiliation{
Centro Multidisciplinar de Astrof\'{\i}sica - CENTRA, 
Departamento de F\'{\i}sica, Instituto Superior T\'ecnico,
Av. Rovisco Pais 1, 1049-001 Lisboa, Portugal
}%

\date{\today}

\begin{abstract}

Gravitational wave solutions to Einstein's equations and their
generation are examined in $D$-dimensional flat spacetimes.  First the
plane wave solutions are analyzed; then the wave generation is
studied with the solution for the metric tensor being obtained with the help
of retarded $D$-dimensional Green's function.  Due to the difficulties 
in handling the wave tails in odd dimensions we concentrate our study in even 
dimensions. We compute the metric
quantities in the wave zone in terms of the energy momentum tensor at
retarded time. Some special cases of interest are
studied: first the slow motion approximation, where the
$D$-dimensional quadrupole formula is deduced. Within the quadrupole
approximation, we consider two cases of interest, a particle in
circular orbit and a particle falling radially into a higher
dimensional Schwarzschild black hole.  Then we turn our attention to
the gravitational radiation emitted during collisions lasting zero
seconds, i.e., hard collisions. We compute the gravitational energy radiated during the
collision of two point particles, in terms of a cutoff frequency. In
the case in which at least one of the particles is a black hole, we
argue this cutoff frequency should be close to the lowest
gravitational quasinormal frequency. In this context, we
compute the scalar quasinormal frequencies of higher dimensional
Schwarzschild black holes.  Finally, as an interesting new application of
this formalism, we compute the gravitational energy release during the
quantum process of black hole pair creation.
These results might be important in light of the recent proposal 
that there may exist extra dimensions in the Universe, one consequence 
of which may be black hole creation at the Large Hadron Collider at CERN.

\end{abstract}

\pacs{04.25.Nx, 04.30.Db, 04.30.-w, 11.10.Kk}

\maketitle
\newpage
\section{Introduction}

One expects to finally detect gravitational waves in the forthcoming
years. If this happens, and if the observed waveforms match the
predicted templates, General Relativity will have pass a crucial test.
Moreover, if one manages to cleanly separate gravitational waveforms,
we will open a new and exciting window to the Universe, a window from
which one can look directly into the heart of matter, as gravitational
waves are weakly scattered by matter.  A
big effort has been spent in the last years trying to build
gravitational wave detectors, and a new era will begin with
gravitational wave astronomy \cite{schutz1,hughes}.  What makes
gravitational wave astronomy attractive, the weakness with which
gravitational waves are scattered by matter, is also the major source
of technical difficulties when trying to develop an apparatus which
interacts with them. Nevertheless, some of these highly non-trivial
technical difficulties have been surmounted, and we have detectors
already operating \cite{geo,ligo,virgo}. Another effort is being
dedicated by theoreticians trying to obtain accurate templates for the
various physical processes that may give rise to the waves impinging
on the detector. We now have a well established theory of wave generation
and propagation, which started with Einstein and his quadrupole formula.  The
quadrupole formula expresses the energy lost to gravitational waves by
a system moving at low velocities, in terms of its energy
content.  The quadrupole formalism is the
most famous example of slow motion techniques to compute wave
generation. All these techniques break Einstein's equations
non-linearity by imposing a power series in some small quantity and
keeping only the lowest or the lowest few order terms. The quadrupole
formalism  starts from a flat background and 
expands the relevant quantities 
in $R/ \lambda$, where $R$ is the size of source and $\lambda$ the 
wavelength of waves.
Perturbation formalisms on the other hand, start from some
non-radiative background, whose metric is known exactly, for example
the Schwarzschild metric, and expand in deviations from that
background metric. For a catalog of the various methods and their
description we refer the reader to the review works by Thorne
\cite{thorne} and Damour \cite{damour}. The necessity to develop all
such methods was driven of course by the lack of exact radiative
solutions to Einstein's equations (although there are some worthy
exceptions, like the C-metric \cite{kinnersley}), 
and by the fact that even nowadays solving the full set of
Einstein's equations numerically is a monumental task, and has been
done only for the more tractable physical situations. All the
existing methods seem to agree with each other when it comes down to
the computation of waveforms and energies radiated during physical
situations, and also agree with the few available results from a fully
numerical evolution of Einstein's equations.

In this work we extend some of these
results to higher dimensional spacetimes.
There are several reasons why one should now try to do it. 
It seems impossible to
formulate in four dimensions a consistent theory which unifies gravity
with the other forces in nature. Thus, most efforts in this direction
have considered a higher dimensional arena for our universe, one
example being string theories which have recently made some remarkable
achievements. Moreover, recent investigations \cite{hamed} propose the
existence of extra dimensions in our Universe in order to solve the
hierarchy problem, i.e., the huge difference between the electroweak
and the Planck scale, $m_{\rm EW}/M_{\rm Pl}\sim 10^{-17}$. 
The fields of standard model would inhabit a 4-dimensional
sub-manifold, the brane, whereas the gravitational degrees of freedom
would propagate throughout all dimensions. One of the most spectacular
consequences of this scenario would be the production of black holes
at the Large Hadron Collider at CERN \cite{bhprod} (for recent
relevant work related to this topic we refer the reader to
\cite{decay,cardosolemos0,cardosolemos}).
Now, one of the experimental
signatures of black hole production will be a missing energy, perhaps
a large fraction of the center of mass energy \cite{cardosolemos0}. 
This will happen because when the partons collide to form a black
hole, some of the initial energy will be converted to gravitational
waves, and due to the small amplitudes involved, there is no
gravitational wave detector capable of detecting them, so they will
appear as missing. Thus, the collider could in fact indirectly serve
as a gravitational wave detector.  This calls for the calculation of
the energy given away as gravitational waves when two high energy
particles collide to form a black hole, which lives in all the
dimensions. The work done so far on this subject
\cite{eardley,yoshino} in higher dimensions, is mostly geometric, and
generalizes a construction by Penrose to find trapped surfaces on the
union of two shock waves, describing boosted Schwarzschild black
holes. On the other hand, there are clues
\cite{cardosolemos0,cardosolemos} indicating that a formalism
described by Weinberg \cite{weinberg} to compute the gravitational
energy radiated in the collision of two point particles, gives results
correct to a order of magnitude when applied to the collision of two
black holes.  This formalism assumes a hard collision, i.e., a
collision lasting zero seconds.  It would be important to apply this
formalism in higher dimensions, trying to see if there is agreement
between both results. This will be one of the topics discussed in this
paper. The other topic we study in this paper is the quadrupole 
formula in higher dimensions. Due to the difficulties in handling the 
wave tails in odd dimensions we concentrate our study in even 
dimensions.

This paper is organized as follows: In section II we linearize
Einstein's equations in a flat $D$-dimensional background and arrive at
an inhomogeneous wave equation for the metric perturbations. The
source free equations are analyzed in terms of plane waves, and then
the general solution to the homogeneous equation is deduced in terms
of the $D$-dimensional retarded Green's function.  In section III we
compute the $D$-dimensional quadrupole formula (assuming slowly moving
sources), expressing the metric and the radiated energy in terms of
the time-time component of the energy-momentum tensor. We then apply
the quadrupole formula to two cases: a particle in circular motion in a 
generic background, and
a particle falling into a $D$-dimensional Schwarzschild black hole.
In section IV we consider the hard collision between two particles, i.e., 
the collision takes zero seconds, and introduce a cutoff frequency
necessary to have meaningful results. We then apply to the case where 
one of the colliding
particles is a black hole. We propose that this cutoff should be
related to the gravitational quasinormal frequency of the black hole, 
and compute some values of the scalar quasinormal frequencies for
higher dimensional Schwarzschild black holes, expecting that the
gravitational quasinormal frequencies will behave in the same manner.  
Finally, we apply this formalism to
compute the generation of gravitational radiation during black hole
pair creation in four and higher dimensions, a result that has never been 
worked out, even for $D=4$.
In our presentation we shall mostly follow Weinberg's \cite{weinberg}
exposition.

\section{Linearized ${\bm D}$-dimensional Einstein's equations}

Due to the non-linearity of Einstein's equations, the treatment of
the gravitational radiation problem is not an easy one since the
energy-momentum tensor of the gravitational wave contributes to
its own gravitational field. To overcome this difficulty it is a
standard procedure to work only with the weak radiative solution,
in the sense that the energy-momentum content of the gravitational
wave is small enough in order to allow us to neglect its
contribution to its own propagation. This approach is justified in
practice since we expect the detected gravitational radiation to
be of low intensity.
\subsection{The inhomogeneous wave equation}
We begin this subsection by introducing the general background
formalism (whose details can be found, e.g., in \cite{weinberg}) that
will be needed in later sections. Then we obtain the linearized
inhomogeneous wave equation.

Greek indices vary as $0,1,\cdots, D-1$ and latin indices as
$1,\cdots, D-1$ and our units are such that $c \equiv 1$.
We work on a $D$-dimensional spacetime described by a
metric $g_{\mu\nu}$ that approaches asymptotically the
$D$-dimensional Minkowski metric $\eta_{\mu\nu}={\rm
diag}(-1,+1,\cdots,+1)$, and thus we can write
\begin{equation}
g_{\mu\nu}=\eta_{\mu\nu}+h_{\mu\nu}\,
\hspace{1cm}\mu,\nu=0,1,\cdots,D-1 \,,
 \label{g}
\end{equation}
where $h_{\mu\nu}$ is small, i.e., $|h_{\mu\nu}|<<1$, so that it
represents small corrections to the flat background.
The exact Einstein field equations,
 $G_{\mu\nu}=8\pi {\cal G} T_{\mu\nu}$ (with ${\cal G}$ being the usual Newton
constant), can then be written as
\begin{equation}
R^{(1)}_{\;\;\;\;\mu\nu}-\frac{1}{2}\eta_{\mu\nu}
 R^{(1)\,\alpha}_{\;\;\;\;\;\;\;\alpha}=8\pi {\cal G}\,\, \tau_{\mu\nu} \,,
 \label{R1}
\end{equation}
with
\begin{equation}
\tau^{\mu\nu} \equiv \eta^{\mu \alpha} \eta^{\nu\beta}
\,(T_{\alpha\beta}+t_{\alpha\beta}).
 \label{tau}
\end{equation}
Here $R^{(1)}_{\;\;\;\;\mu\nu}$ is the part of the Ricci tensor
linear in $h_{\mu\nu}$,
 $R^{(1)\,\alpha}_{\;\;\;\;\;\;\;\alpha}=\eta^{\alpha\beta}
 R^{(1)}_{\;\;\;\;\beta\alpha}$,
and $\tau_{\mu\nu}$ is the effective energy-momentum tensor, containing
contributions from $T_{\mu\nu}$, the energy-momentum tensor of the matter source,
and $t_{\mu\nu}$ which represents the gravitational contribution.
The pseudo-tensor $t_{\mu\nu}$ contains the difference between
the exact Ricci terms and the Ricci terms linear in $h_{\mu\nu}$,
\begin{equation}
t_{\mu\nu}=\frac{1}{8\pi {\cal G}}\left [ R_{\mu\nu}
-\frac{1}{2}g_{\mu\nu} R^{\alpha}_{\;\;\;\alpha}
 -R^{(1)}_{\;\;\;\;\mu\nu}+\frac{1}{2}\eta_{\mu\nu}
 R^{(1)\,\alpha}_{\;\;\;\;\;\;\;\alpha} \right ] \,.
 \label{t}
\end{equation}
The Bianchi identities imply that $\tau_{\mu\nu}$ is locally conserved,
\begin{equation}
\partial_{\mu} \tau^{\mu\nu}=0 \,.
 \label{constau}
\end{equation}
Introducing the cartesian coordinates $x^{\alpha}=(t,{\bf x})$ with ${\bf
x}=x^i$, and considering a $D-1$ volume $V$ with a boundary spacelike
surface $S$ with dimension $D-2$ whose unit exterior normal is ${\bf
n}$, eq. (\ref{constau}) yields
\begin{equation}
\frac{d}{dt}\int_{V} d^{D-1}{\bf x} \;\tau^{0\nu}
=-\int_{S}
 d^{D-2}{\bf x} \;n_i \tau^{i\nu} \,.
 \label{constau1}
\end{equation}
This means that one may interpret
\begin{equation}
p^{\nu} \equiv \int_{V} d^{D-1}{\bf x} \;\tau^{0\nu}
 \label{p}
\end{equation}
as the total energy-momentum (pseudo)vector of the system, including
matter and gravitation, and $\tau^{i\nu}$ as the corresponding flux.
Since the matter contribution is contained in $t^{\mu\nu}$,
the flux of gravitational radiation is
\begin{equation}
{\rm Flux}=\int_{S} d^{D-2}{\bf x} \;n_i t^{i \nu} \,.
 \label{fluxgrav}
\end{equation}

In this context of linearized general
relativity, we neglect terms of order higher than the first in
$h_{\mu\nu}$ and all the indices are raised and lowered using
$\eta^{\mu\nu}$. We also neglect the contribution of the
gravitational energy-momentum tensor $t_{\mu\nu}$ (i.e.,
$|t_{\mu\nu}|<<|T_{\mu\nu}|$) since from (\ref{t}) we see that
$t_{\mu\nu}$ is of higher order in $h_{\mu\nu}$. Then, the
conservation equations (\ref{constau}) yield
\begin{equation}
\partial_{\mu}T^{\mu\nu}=0 \,.
 \label{consT}
\end{equation}
In this setting and choosing the convenient coordinate system
that obeys the harmonic (also called Lorentz) gauge conditions,
\begin{equation}
2\partial_{\mu}h^{\mu}_{\;\;\;\nu}=\partial_{\nu}h^{\alpha}_{\;\;\;\alpha}
 \label{hargauge}
\end{equation}
(where $\partial_{\mu}=\partial/\partial x^{\mu}$), the first
order Einstein field equations (\ref{R1}) yield
\begin{equation}
\square h_{\mu\nu}=-16\pi {\cal G} S_{\mu\nu} \,,
 \label{ineq}
\end{equation}
\begin{equation}
S_{\mu\nu}=T_{\mu\nu}-\frac{1}{D-2}\,\eta_{\mu\nu}\,T^{\alpha}_{\;\;\;\alpha}
\,,
 \label{S}
\end{equation}
where $\square=\eta^{\mu\nu}\partial_{\mu}\partial_{\nu}$ is the
$D$-dimensional Laplacian, and $S_{\mu\nu}$ will be called the
modified energy-momentum tensor of the matter source. Eqs.
(\ref{ineq}) and (\ref{S}) subject to (\ref{hargauge}) allow us to
find the gravitational radiation produced by a matter source
$S_{\mu\nu}$.
\subsection{The plane wave solutions}

In vacuum, the linearized equations for the gravitational field are
$R^{(1)}_{\;\;\;\;\mu\nu}=0$ or, equivalently, the homogeneous
equations $\square h_{\mu\nu}=0$, subjected to the harmonic gauge
conditions (\ref{hargauge}). The solutions of these equations, the
plane wave solutions, are important since the general solutions of the
inhomogeneous equations (\ref{hargauge}) and (\ref{ineq}) approach the
plane wave solutions at large distances from the source.  Setting
$k_{\alpha}=(-\omega,{\bf k})$ with $\omega$ and ${\bf k}$ being
respectively the frequency and wave vector, the plane wave solutions
can be written as a linear superposition of solutions of the kind
\begin{equation}
 h_{\mu\nu}(t,{\bf x})=e_{\mu\nu}\,e^{i k_{\alpha} x^{\alpha}}+
 e_{\mu\nu}^*\,e^{-i k_{\alpha} x^{\alpha}} \,,
  \label{homsol}
\end{equation}
where $e_{\mu\nu}=e_{\nu\mu}$ is called the polarization tensor
and $^*$ means the complex conjugate. These solutions satisfy eq.
(\ref{ineq}) with $S_{\mu\nu}=0$ if $k_{\alpha} k^{\alpha}=0$, and
obey the harmonic gauge conditions (\ref{hargauge}) if
$2\,k_{\mu}\,e^{\mu}_{\;\;\;\nu}=k_{\nu}\,e^{\mu}_{\;\;\;\mu}$.

An important issue that must be addressed is the number of
different polarizations that a gravitational wave in $D$
dimensions can have. The polarization tensor $e_{\mu\nu}$, being
symmetric, has in general $D(D+1)/2$ independent components.
However, these components are subjected to the $D$ harmonic gauge
conditions that reduce by $D$ the number of independent
components. In addition, under the infinitesimal change of
coordinates $x'^{\mu}=x^{\mu}+\xi^{\mu}(x)$, the
polarization tensor transforms into
$e'_{\mu\nu}=e_{\mu\nu}-\partial_{\nu}\xi_{\mu}
-\partial_{\mu}\xi_{\nu}$. Now, $e'_{\mu\nu}$ and
$e_{\mu\nu}$ describe the same physical system for arbitrary
values of the $D$ parameters $\xi^{\mu}(x)$. Therefore, the number
of independent components of $e_{\mu\nu}$, i.e., the number of
polarization states of a gravitational wave in $D$ dimensions is
$D(D+1)/2-D-D=D(D-3)/2$. From this computation we can also see
that gravitational waves are present only when $D>3$. Therefore,
from now on we assume ${\rm D}>3$ whenever we refer to $D$. In what concerns
the helicity of the gravitational waves, for arbitrary $D$ the
gravitons are always spin $2$ particles.

To end this subsection on gravitational plane wave solutions, we
present the average gravitational energy-momentum tensor of a
plane wave, a quantity that will be needed later. Notice that
in vacuum, since the matter contribution is zero
($T_{\mu\nu}=0$), we cannot neglect the contribution of the
gravitational energy-momentum tensor $t_{\mu\nu}$. From eq.
(\ref{t}), and neglecting terms of order higher than $h^2$, the
gravitational energy-momentum tensor of a plane wave is given by
\begin{equation}
t_{\mu\nu}\simeq \frac{1}{8\pi {\cal G} }\left [
R^{(2)}_{\;\;\;\;\mu\nu}-\frac{1}{2}\eta_{\mu\nu}
 R^{(2)\,\alpha}_{\;\;\;\;\;\;\;\alpha} \right ] \,,
 \label{tplane}
\end{equation}
and through a straightforward calculation (see e.g.
\cite{weinberg} for details) we get the average gravitational
energy-momentum tensor of a plane wave,
\begin{equation}
\langle t_{\mu\nu} \rangle=\frac{k_{\mu}k_{\nu}}{16\pi {\cal G}}\left [
e^{\alpha\beta}e_{\alpha\beta}^*
-\frac{1}{2}|e^{\alpha}_{\;\;\;\alpha}|^2\right ] \,.
 \label{taverage}
\end{equation}

\subsection{The $\bm D$-dimensional retarded Green's function}
The general solution to the inhomogeneous differential 
equation (\ref{ineq}) may be found in the usual way in terms of
a Green's function as 
\begin{equation}
h_{\mu\nu}(t,{\bf x})=-16 \pi {\cal G}\int dt'\int d^{D-1}{\bf x'} 
S_{\mu \nu}(t',{\bf x'})
G(t-t',{\bf x - x'})+{\rm homogeneous \,solutions}\,,
\label{insol}
\end{equation}
where the Green's function $G(t-t',{\bf x - x'})$ satisfies
\begin{equation}
\eta^{\mu\nu}\partial_{\mu}\partial_{\nu} 
G(t-t',{\bf x - x'})=\delta(t-t')\delta({\bf x - x'})\,,
\label{greendef}
\end{equation}
where $\delta(z)$ is the Dirac delta function.
In the momentum representation this reads 
\begin{equation}
G(t,{\bf x})=-\frac{1}{(2\pi)^{D}} \int d^{D-1} {\bf k} 
e^{i {\bf k}\cdot{\bf x}}\int d\omega \frac{e^{-i\omega t}}{\omega^2-k^2}\,,
\label{greenmomentum}
\end{equation}
where $k^2=k_{1}^2+k_{2}^2+...+k_{D-1}^2$.
To evaluate this, it is convenient to perform the $k$-integral by
using spherical coordinates in the ($D-1$)-dimensional $k$-space.  The
required transformation, along with some useful formulas which shall
be used later on, is given in Appendix \ref{apendice}.  The result for the retarded
Green's function in those spherical coordinates is
\begin{equation}
G^{\rm ret}(t,{\bf x})=-\frac{\Theta(t)}{(2\pi)^{(D-1)/2}}\times\frac{1}{r^{(D-3)/2}}
\int k^{(D-3)/2} J_{(D-3)/2}(kr)\sin(kt)dk \,,
\label{greenretgeral}
\end{equation}
where $r^2=x_{1}^2+x_{2}^2+...+x_{D-1}^2$, and  $\Theta(t)$ is the Heaviside 
function defined as
\begin{equation}
\Theta(t)=\left\{ \begin{array}{ll}
             1   & \mbox{if $t>0$}\\
             0    & \mbox{if $t<0$}\,.
\end{array}\right.
\label{Heaviside}
\end{equation}
The function $J_{({\rm D}-3)/2}(kr)$ is a Bessel function
\cite{watson,stegun}.  The structure of the retarded Green's function
will depend on the parity of $D$, as we shall see. This dependence on
the parity, which implies major differences between even and odd
spacetime dimensions, is connected to the structure of the Bessel
function. For even $D$, the index of the Bessel function is
semi-integer and then the Bessel function is expressible in terms of
elementary functions, while for odd $D$ this does not happen. A
concise explanation of the difference between retarded Green's
function in even and odd $D$, and the physical consequences that
entails is presented in \cite{courant} (see also \cite{hadamard,barrow,galtsov}).
 A complete derivation of the Green's function in
higher dimensional spaces may be found in Hassani \cite{hassani}.
The result is
\begin{equation}
G^{\rm ret}(t,{\bf x})=\frac{1}{4\pi}\left[-\frac{\partial}{2\pi r \partial r} \right]^{(D-4)/2}
\left[\frac{\delta(t-r)}{r}\right]\,,\,\,\,\,D \,\,{\rm even}.
\label{greenfinaleven}
\end{equation}
\begin{equation}
G^{\rm ret}(t,{\bf x})=\frac{\Theta(t)}{2\pi}\left[-\frac{\partial}{2\pi r \partial r} \right]^{(D-3)/2}
\left[\frac{1}{\sqrt{t^2-r^2}}\right]\,,\,\,\,\,D \,\,{\rm odd}.
\label{greenfinalodd}
\end{equation}
It is sometimes convenient to work with the Fourier transform (in the time
coordinate) of the Green's function. One finds \cite{hassani} an
analytical result independent of the parity of $D$
\begin{equation}
G^{\rm ret}(\omega,{\bf x})=\frac{i^D \pi}{2(2\pi)^{(D-1)/2}} 
\left(\frac{\omega}{r}\right)^{(D-3)/2} H^1_{(D-3)/2}(\omega r)\,,
\label{greenfinalevenfourier}
\end{equation}
where $H^1_{\nu}(z)$ is a modified Bessel function
\cite{stegun,watson}.  Of course, the different structure of the
Green's function for different $D$ is again embodied in these Bessel
functions.  Equations (\ref{greenfinaleven}) and
(\ref{greenfinalevenfourier}), are one of the most important results
we shall use in this paper.  For $D=4$ (\ref{greenfinaleven})
obviously reproduce well known results \cite{hassani}.  Now, one sees
from eq. (\ref{greenfinaleven}) that although there are delta function
derivatives on the even-$D$ Green's function, the localization of the
Green's function on the light cone is preserved.
However, eq. (\ref{greenfinalodd}) tells us that the retarded Green's
function for odd dimensions is non-zero inside the light
cone. The consequence, as has been emphasized by different authors
\cite{courant,galtsov, kazinski}, is that for odd $D$ the Huygens principle
does not hold: the fact that the retarded Green's function support
extends to the interior of the light cone implies the appearance
of radiative tails in (\ref{insol}). 
In other words, we still have a propagation
phenomenum for the wave equation in odd dimensional spacetimes, in so
far as a localized initial state requires a certain time to reach a
point in space. Huygens principle no longer holds, because
the effect of the initial state is not sharply limited in time: once
the signal has reached a point in space, it persists there
indefinitely as a reverberation.

This fact coupled to the analytic structure of the Green's function in
odd dimensions make it hard to get a grip on radiation generation in
odd dimensional spacetimes. Therefore, from now on we shall focus on
even dimensions, for which the retarded Green's function is given by
eq. (\ref{greenfinaleven}).

\subsection{The even ${\bm D}$-dimensional retarded solution in the wave zone}
The retarded solution for the metric perturbation $h_{\mu\nu}$,
obtained by using the retarded Green's function (\ref{greenfinaleven})
and discarding the homogeneous solution in (\ref{insol})
will be given by
\begin{equation}
h_{\mu\nu}(t,{\bf x})= 16 \pi {\cal G}\int dt'\int d^{D-1}{\bf x'} 
S_{\mu \nu}(t',{\bf x'})
G^{\rm ret}(t-t', {\bf x - x'})\,,
\label{retsol}
\end{equation}
with $G^{\rm ret}(t-t',{\bf x - x'})$ as in eq. (\ref{greenfinaleven}).
For $D=4$ for example one has
\begin{equation}
G^{\rm ret}(t,{\bf x})=\frac{1}{4\pi}\frac{\delta(t-r)}{r}\,\,,\,\,\,\,\,D=4\,,
\label{green4}
\end{equation}
which is the well known result.
For $D=6$, we have
\begin{equation}
G^{\rm ret}(t,{\bf x})=\frac{1}{8\pi^2}\left(\frac{\delta'(t-r)}{r^2}+\frac{\delta(t-r)}{r^3}\right)
\,\,,\,\,\,\,\,D=6\,,
\label{green6}
\end{equation}
where the $\delta'(t-r)$ means derivative of the Dirac delta function with respect to its argument.
For $D=8$, we have
\begin{equation}
G^{\rm ret}(t,{\bf x})=\frac{1}{16\pi^3}\left(\frac{\delta''(t-r)}{r^3}+
3\frac{\delta'(t-r)}{r^4}+3\frac{\delta(t-r)}{r^5}\right)
\,\,,\,\,\,\,\,D=8\,.
\label{green8}
\end{equation}
We see that in general even-$D$ dimensions the Green's function
consists of inverse integer powers in r, spanning all values between
$\frac{1}{r^{(D-2)/2}}$ and $\frac{1}{r^{D-3}}$, including these ones.
Now, the retarded solution is given by eq. (\ref{retsol}) as a product
of the Green's function times the modified energy-momentum tensor
$S_{\mu \nu}$. The net result of having derivatives on the delta
functions is to transfer these derivatives to the energy-momentum
tensor as time derivatives (this can be seen by integrating
(\ref{retsol}) by parts in the $t$-integral).

A close inspection then shows that the retarded field possesses a kind
of peeling property in that it consists of terms with different
fall off at infinity. Explicitly, this means that the retarded field
will consist of a sum of terms possessing all integer inverse powers
in $r$ between $\frac{D-2}{2}$ and $D-3$.  The term that dies off more
quickly at infinity is the $\frac{1}{r^{D-3}}$, typically a
static term, since it comes from the Laplacian.  As a matter of fact
this term was already observed in the higher dimensional black hole
by Tangherlini \cite{tangherlini} (see also Myers and Perry 
\cite{myersperry}). 
We will see that the
term falling more slowly, the one that goes like
$\frac{1}{r^{(D-2)/2}}$, gives rise to gravitational radiation. It is 
well defined, in the sense that the power crossing
sufficiently large hyperspheres with different radius is the same, because
the volume element goes as $r^{D-2}$ and the energy as $|h|^2
\sim \frac{1}{r^{D-2}}$.

In radiation problems, one is interested in finding out the field at
large distances from the source, $r>>\lambda$, where $\lambda$ is the 
wavelength of the waves, and also much larger
than the source's dimensions $R$. This is defined as the wave zone.  In
the wave zone, one may neglect all terms in the Green's function that
decay faster than $\frac{1}{r^{(D-2)/2}}$. So, in the wave zone, we
find
\begin{equation}
h_{\mu\nu}(t,{\bf x})= -8 \pi {\cal G} \frac{1}{(2\pi r)^{(D-2)/2}}
\partial_{t}^{(\frac{D-4}{2})}\left[ \int d^{D-1}{\bf x'} 
 S_{\mu \nu}(t-|{\bf x-x'}|,{\bf x'})\right]\,,
\label{retsolwavezone}
\end{equation}
where $\partial_{t}^{(\frac{D-4}{2})}$ stands for the $\frac{D-4}{2}$th derivative
with respect to time.
For $D=4$ eq. (\ref{retsolwavezone}) yields the standard result
\cite{weinberg}:
\begin{equation}
h_{\mu\nu}(t,{\bf x})= -\frac{4 {\cal G}}{r} \int d^{D-1}{\bf x'} 
 S_{\mu \nu}(t-|{\bf x-x'}|,{\bf x'})\,,\,\,\,\,\,D=4.
\label{retsolwavezoned4}
\end{equation}
To find the Fourier transform of the metric, one uses the
representation (\ref{greenfinalevenfourier}) for the Green's
function. Now, in the wave zone, the Green's function may be
simplified using the asymptotic expansion for the Bessel function
\cite{stegun}
\begin{equation}
H^1_{(D-3)/2}(\omega r) \sim \sqrt{\frac{2}{\pi(\omega r)}} 
e^{i\left[\omega r-\frac{\pi}{4}(D-2)\right]}\,\,,\,\,\,\,\,
\omega r \rightarrow \infty.
\label{asymbessel}
\end{equation}
This yields 
\begin{equation}
h_{\mu\nu}(\omega,{\bf x})= -\frac{8 \pi {\cal G}}{(2\pi r)^{(D-2)/2}}\omega^{(D-4)/2}e^{i\omega r}
\int d^{D-1}{\bf x'} S_{\mu \nu}(\omega,{\bf x'})\,.
\label{retsolwavezonefourier}
\end{equation}
This could also have been arrived at directly from
(\ref{retsolwavezone}), using the rule time derivative $\rightarrow
-i\omega $ for Fourier transforms.  Equations (\ref{retsolwavezone})
and (\ref{retsolwavezonefourier}) are one of the most important
results derived in this paper, and will be the basis for all the
subsequent section. Similar equations, but not as general as the ones 
presented here, were given by Chen, 
Li and Lin \cite{lin} in the context of gravitational radiation by 
a rolling tachyon. 
 
To get the energy spectrum, we use (\ref{S})
yielding
\begin{equation}
\frac{d^2E}{d\omega d\Omega}= 2 {\cal G} \frac{\omega^{D-2}}{(2\pi)^{D-4}} 
\left( T^{\mu\nu}(\omega,{\bf k})T_{\mu\nu}^*(\omega,{\bf k})-\frac{1}{D-2}
|T^{\lambda}_{\:\:\:\lambda}(\omega, {\bf k})|^2\right)\,.
\label{powerwavezone}
\end{equation}

\section{The even $\bm D$-dimensional quadrupole formula}
\subsection{Derivation of the even $\bm D$-dimensional quadrupole formula}
When the velocities of the sources that generate the gravitational
waves are small, it is sufficient to know the $T^{00}$ component of
the gravitational energy-momentum tensor in order to have a good
estimate of the energy they radiate. In this subsection, we will
deduce the $D$-dimensional quadrupole formula and in the next
subsection we will apply it to (1) a particle in circular orbit and
(2) a particle in free fall into a $D$-dimensional Schwarzschild black
hole.

We start by recalling that the Fourier transform of the
energy-momentum tensor is
\begin{equation}
T_{\mu\nu}(\omega,{\bf k})=\int d^{D-1}{\bf x'} e^{-i {\bf k}\cdot
{\bf x'}} \int dt\, e^{i \omega t}\,T^{\mu\nu}(t,{\bf x})+ {\rm c.c.} \,,
\label{fourierT}
\end{equation}
where ${\rm c.c.}$ means the complex conjugate of the preceding term.
Then, the conservation equations (\ref{consT}) for
$T^{\mu\nu}(t,{\bf x})$ applied to eq. (\ref{fourierT}) yield
$k^{\mu}\,T_{\mu\nu}(\omega,{\bf k})=0$. Using this last result we
obtain $T_{00}(\omega,{\bf k})=\hat{k}^j \,\hat{k}^i
\,T_{ji}(\omega,{\bf k})$ and $T_{0i}(\omega,{\bf k})=-\hat{k}^j\,
T_{ji}(\omega,{\bf k})$, where ${\bf \hat{k}}={\bf k}/\omega$. 
We can then write the energy spectrum, eq.
(\ref{powerwavezone}), as a function only of the spacelike
components of $T^{\mu\nu}(\omega,{\bf k})$,
\begin{equation}
\frac{d^2E}{d\omega d\Omega}= 2 {\cal G}
\frac{\omega^{D-2}}{(2\pi)^{D-4}}\, \Lambda_{ij,\,lm}(\hat{k})\,
T^{*\,ij}(\omega,{\bf k})\, T^{\,ij}(\omega,{\bf k}) \,,
\label{powerwavezone2}
\end{equation}
where
\begin{equation}
 \Lambda_{ij,\,lm}(\hat{k})=\delta_{il}\delta_{jm}
 -2\hat{k}_j \hat{k}_m \delta_{il}
 +\frac{1}{D-2}\left (-\delta_{ij}\delta_{lm}
 + \hat{k}_l \hat{k}_m \delta_{ij}
 +\hat{k}_i \hat{k}_j \delta_{lm}\right )
 +\frac{D-3}{D-2}\hat{k}_i \hat{k}_j \hat{k}_l \hat{k}_m \,.
\label{Lambda}
\end{equation}
At this point, we make a new approximation (in addition to the
wave zone approximation) and assume that $\omega R<<1$, where $R$
is the source's radius. In other words, we assume that the internal 
velocities of the sources are small and thus the source's radius is much smaller
than the characteristic wavelength $\sim 1/\omega$ of the emitted gravitational
waves. Within this approximation, one can set $e^{-i {\bf k}\cdot
{\bf x'}} \sim 1$  in eq. (\ref{fourierT}) (since $R=|{\bf
x'}|_{\rm max}$). Moreover, after a straightforward calculation,
one can also set in eq. (\ref{powerwavezone2}) the approximation
$T^{\,ij}(\omega,{\bf k})\simeq -(\omega^2/2)D_{ij}(\omega)$,
where
\begin{equation}
D_{ij}(\omega)=
 \int d^{D-1} {\bf x}\, x^i \,x^j \,T^{00}(\omega,{\bf x})\,.
 \label{Dij}
\end{equation}
Finally, using
\begin{eqnarray}
\int d\Omega_{D-2}\hat{k}_i
\hat{k}_j=\frac{\Omega_{D-2}}{D-1}\delta_{ij}\,, \nonumber \\
\int d\Omega_{D-2}\hat{k}_i \hat{k}_j \hat{k}_l \hat{k}_m=
 \frac{3\Omega_{D-2}}{D^2-1} (\delta_{ij}\delta_{lm}
  +\delta_{il}\delta_{jm}+\delta_{im}\delta_{jl})\,,
\label{intk}
\end{eqnarray}
where $\Omega_{D-2}$ is the $(D-2)$-dimensional solid angle
defined in (\ref{integratedsolidangle}), we obtain the
$D$-dimensional quadrupole formula
\begin{equation}
\frac{dE}{d\omega}=\frac{2^{2-D}\pi^{-(D-5)/2}{\cal G}\,(D-3)D}
{\Gamma[(D-1)/2](D^2-1)(D-2)}\, \omega^{D+2} {\biggl [}
(D-1)D^*_{ij}(\omega)D_{ij}(\omega)-|D_{ii}(\omega)|^2 {\biggr ]}
 \,,
\label{quadw}
\end{equation}
where the Gamma function $\Gamma[z]$ is defined in Appendix
\ref{apendice}.  As the dimension $D$ grows it is seen that the rate
of gravitational energy radiated increases as $\omega^{D+2}$.
Sometimes it will be more useful to have the time rate of emitted
energy
\begin{equation}
\frac{dE}{dt}=\frac{2^{2-D}\pi^{-(D-5)/2} {\cal G}\,(D-3)D}
{\Gamma[(D-1)/2](D^2-1)(D-2)}\,{\biggl [}
(D-1)\partial_t^{(D+2)/2}D^*_{ij}(t)\partial_t^{(D+2)/2}D_{ij}(t)-
|\partial_t^{(D+2)/2}D_{ii}(t)|^2 {\biggr ]}
 \,.
\label{quad}
\end{equation}
For $D=4$, eq. (\ref{quad}) yields the well known result \cite{weinberg}
\begin{equation}
\frac{dE}{dt}=\frac{{\cal G}}{5}\, {\biggl [}
\partial_t^{3}D^*_{ij}(t)\partial_t^{3}D_{ij}(t)-
\frac{1}{3}|\partial_t^{3}D_{ii}(t)|^2
{\biggr ]}
 \,.
\label{quad4}
\end{equation}

\subsection{Applications of the quadrupole formula: test particles in a background geometry}
The quadrupole formula has been used successfully
in almost all kind of problems involving gravitational wave
generation. By successful we mean that it agrees with other
more accurate methods. Its simplicity and the fact that it gives
results correct to within a few percent, 
makes it an invaluable tool in estimating
gravitational radiation emission. We shall in the following present
two important examples of the application of the quadrupole formula.
\subsubsection{A particle in circular orbit}
The radiation generated by particles in circular motion was perhaps
the first situation to be considered in the analysis of gravitational
wave generation. For orbits with low frequency, the quadrupole formula
yields excellent results. As expected it is difficult to find in
nature a system with perfect circular orbits, they will in general be
elliptic. In this case the agreement is also remarkable, and one finds
that the quadrupole formalism can account with precision for the
increase in period of the pulsar PSR 1913$+$16, due to gravitational
wave emission \cite{pulsar}. In four dimensions the full treatment of
elliptic orbital motion is discussed by Peters \cite{peters}.  In
dimensions higher than four, it has been shown \cite{tangherlini} that
there are no stable geodesic circular orbits, and so geodesic circular
motion is not as interesting for higher $D$. For this reason, and also
because we only want to put in evidence the differences that arise in
gravitational wave emission as one varies the spacetime dimension $D$, we
will just analyze the simple circular, not necessarily geodesic
motion, to see whether the results are non-trivially changed as one
increases $D$.  Consider then two bodies of equal mass $m$ in circular
orbits a distance $l$ apart.  Suppose they revolve around the center
of mass, which is at $l/2$ from both masses, and that they orbit with
frequency $\omega$ in the $x-y$ plane.  A simple calculation
\cite{peters,schutzlivro} yields
\begin{eqnarray}
D_{xx}=\frac{ml^2}{4}\cos({2\omega t}) \,+\,{\rm const}\,\,, \\
D_{yy}=-D_{xx}\,\,, \\
D_{xy}=\frac{ml^2}{4}\sin({2\omega t}) \,+\,{\rm const}\,\,,
\label{moments}
\end{eqnarray}
independently of the dimension in which they are imbedded and with all
other components being zero.  We therefore get from eq. (\ref{quad})

\begin{equation}
\frac{dE}{dt}=\frac{2{\cal G}D(D-3)}{\pi^{(D-5)/2}\Gamma[(D-1)/2](D+1)(D-2)}m^2 l^4
 \omega^{D+2}.
\label{totalenecircular}
\end{equation}
For $D=4$ one gets
\begin{equation}
\frac{dE}{dt}=\frac{8{\cal G}}{5}m^2 l^4 \omega^{6}\,,
\label{totalenecircularD4}
\end{equation}
which agrees with known results \cite{peters,schutzlivro}.
Eq. (\ref{totalenecircular}) is telling us that as one climbs up
in dimension number $D$, the frequency effects gets more pronounced.

\subsubsection{A particle falling radially into a higher dimensional
Schwarzschild black hole}
As yet another example of the use of the quadrupole formula
eq. (\ref{quad}) we now calculate the energy given away as
gravitational waves when a point particle, with mass $m$ falls into a
$D$-dimensional Schwarzschild black hole, a metric first given in
\cite{tangherlini}.  Historically, the case of a particle falling into
a $D=4$ Schwarzschild black hole was one of the first to be studied
\cite{zerilli,davis} in connection with gravitational wave generation,
and later served as a model calculation when one wanted to evolve
Einstein's equations fully numerically \cite{smarr,gleiser}.  This
process was first studied \cite{davis} by solving numerically
Zerilli's \cite{zerilli} wave equation for a particle at rest at
infinity and then falling into a Schwarzschild black hole. Davis et al
\cite{davis} found numerically that the amount of energy radiated to
infinity as gravitational waves was $\Delta E_{\rm num} =0.01
\frac{m^2}{M}$, where $m$ is the mass of the particle falling in and
$M$ is the mass of the black hole.
It is found that the $D=4$ quadrupole formula yields \cite{quem}
$\Delta E_{\rm quad} =0.019 \frac{m^2}{M}$, so it is of the order of
magnitude as that given by fully relativistic numerical results.
Despite the fact that the quadrupole formula fails somewhere near the
black hole (the motion is not slow, and the background is certainly
not flat), it looks like one can get an idea of how much radiation
will be released with the help of this formula.  Based on this good
agreement, we shall now consider this process but for higher
dimensional spacetimes.  The metric for the $D$-dimensional
Schwarzschild black hole in ($t,r,\theta_1,\theta_2,..,\theta_{D-2}$)
coordinates (see Appendix \ref{apendice}) is
\begin{equation}
ds^2= -\left(1-\frac{16\pi{\cal G} M}{(D-2)\Omega_{D-2}}\frac{1}{r^{D-3}}\right)dt^2+
\left(1-\frac{16\pi{\cal G} M}{(D-2)\Omega_{D-2}}\frac{1}{r^{D-3}}\right)^{-1}dr^2
+r^{D-2}d\Omega_{D-2}^2.
\label{metricmyersperry} 
\end{equation}
Consider a particle falling along a radial geodesic, and at rest at infinity.
Then, the geodesic equations give
\begin{equation}
\frac{dr}{dt} \sim \frac{16\pi{\cal G} M}{(D-2)\Omega_{D-2}}\frac{1}{r^{D-3}}\,,
\label{geodesic}
\end{equation}
where we make the flat space approximation $t=\tau$.  We then have, in
these coordinates, $D_{11}=r^2$, and all other components vanish.
From (\ref{quad}) we get the energy radiated per second, which yields
\begin{equation}
\frac{dE}{dt}= \frac{2^{2-D}\pi^{-(D-5)/2}{\cal G}\,(D-3)}
{\Gamma[(D-1)/2](D^2-1)}D|\partial_{t}^{(\frac{D+2}{2})}
D_{11}|^2\,,
\label{enerpersecinfall}
\end{equation}
We can perform the derivatives and integrate to get the total energy
radiated.  There is a slight problem though, where do we stop the
integration?  The expression for the energy diverges at $r=0$ but this
is no problem, as we know that as the particle approaches the horizon,
the radiation will be infinitely red-shifted. Moreover, the standard
picture \cite{quem} is that of a particle falling in, and in the last
stages being frozen near the horizon.  With this in mind we integrate
from $r=\infty$ to some point near the horizon, say $r=b\times r_+$,
where $r_+$ is the horizon radius and $b$ is some number larger than
unit, and we get
\begin{equation}
\Delta E=
A \frac{D(D-2)\pi}{2^{2D-4}} \times b^{(9-D^2)/2} \times \frac{m^2}{M}\,,
\label{totalenergyinfall}
\end{equation}
where
\begin{equation}
A=\frac{(3-D)^2(5-D)^2(7-3D)^2(8-4D)^2(9-5D)^2...
(D/2+4-D^2/2)^2}{\Gamma[(D-1)/2]^2(D-1)(D+1)(D+3)}
\label{A}
\end{equation}
To understand the effect of both the dimension number $D$ and the
parameter $b$ on the total energy radiated according to the quadrupole
formula, we list in Table 1 some values $\Delta E$ for different
dimensions, and $b$ between $1$ and $1.3$.

\vskip 1mm
\begin{table}
\caption{\label{tab:zzz} The energy radiated by a particle falling
from rest into a higher dimensional Schwarzschild black hole, as a
function of dimension.  The integration is stopped at $b\times r_+$
where $r_+$ is the horizon radius.}
\begin{ruledtabular}
\begin{tabular}{llll}  \hline
\multicolumn{1}{c}{} &
\multicolumn{3}{c}{ $\Delta E \times \frac{M}{m^2}$}\\ \hline
$D$ & $b=1$:  &     $b=1.2$:   & $b=1.3$:\\ \hline
4   &  0.019  &  0.01 &  0.0076 \\ \hline 
6   &  0.576  &  0.05 &  0.0167 \\ \hline 
8   &  180    &  1.19 &  0.13   \\ \hline 
10  &  24567  &  6.13 &  0.16   \\ \hline 
12  &$3.3\times 10^6$ & 14.77 & 0.0665 \\ \hline 
\end{tabular}
\end{ruledtabular}
\end{table}
\vskip 1mm
The parameter $b$ is in fact a measure of our ignorance of what goes
on near the black hole horizon, so if the energy radiated doesn't vary
much with $b$ it means that our lack of knowledge doesn't affect the
results very much.  For $D=4$ that happens indeed. Putting $b=1$ gives
only an energy $2.6$ times larger than with $b=1.3$, and still very
close to the fully relativistic numerical result of $0.01
\frac{m^2}{M}$.  However as we increase $D$, the effect of $b$
increases dramatically. For $D=12$ for example, we can see that a
change in $b$ from $1$ to $1.3$ gives a corresponding change in
$\Delta E$ of $3\times 10^6$ to $0.0665$. This is $8$ orders of
magnitude lower!  Since there is as yet no 
Regge-Wheeler-Zerilli \cite{zerilli,regge} wavefunction for higher
dimensional Schwarzschild black holes, there are no fully
relativistic numerical results to compare our results with. 
Thus $D=4$ is just the perfect dimension to predict, through the
quadrupole formula, the gravitational energy coming from collisions
between particles and black holes, or between small and massive black
holes.  It is not a problem related to the quadrupole formalism, but
rather one related to $D$. A small change in parameters translates
itself, for high $D$, in a large variation in the final result. Thus,
as the dimension $D$ grows, the knowledge of the cutoff radius
$b\times r_+$ becomes essential to compute accurately the energy
released.

\section{Instantaneous Collisions in Even $\bm D$-Dimensions}
In general, whenever two bodies collide or scatter there will be
gravitational energy released due to the changes in momentum involved
in the process. If the collision is hard meaning that the incoming
and outgoing trajectories have constant velocities, there is a method
first envisaged by Weinberg \cite{weinberg,wein1}, later explored
in \cite{wein2} by Smarr to compute exactly the metric perturbation and energy
released.The method is valid for arbitrary velocities 
(one will still be working in the linear approximation, so energies have to be
low). Basically, it assumes a collision lasting for zero
seconds. It was found that in this case the resulting spectra were flat,
precisely what one would expect based on one's experience with
electromagnetism \cite{jackson}, and so to give a meaning to the total
energy, a cutoff frequency is needed. This cutoff frequency depends
upon some physical cutoff in the particular problem.
We shall now generalize this construction for arbitrary dimensions.
\subsection{Derivation of the Radiation Formula 
in Terms of a Cutoff for a Head-on Collision}
\label{HeadonCollision}
Consider therefore a system of freely moving particles with
$D$-momenta $P_{i}^{\mu}$, energies $E_i$ and ($D-1$)-velocities
${\bf v}$, which due to the collision change abruptly at $t=0$, to
corresponding primed quantities. For such a system, the
energy-momentum tensor is
\begin{equation}
T^{\mu\nu}(t, {\bf v})=  
\sum \frac{P_{i}^{\mu}P_{i}^{\nu}}{E_i} 
\delta^{D-1}({\bf x}-{\bf v}t)\Theta(-t)+
\frac{{P'}_{i}^{\mu}{P'}_{i}^{\nu}}{E'_i} 
\delta^{D-1}({\bf x'}-{\bf v'}t)\Theta(t)\,,
\label{enmomtenpointpctles}
\end{equation}
from which, using eqs. (\ref{retsolwavezonefourier}) and
(\ref{powerwavezone}) one can get the quantities $h_{\mu\nu}$ and also
the radiation emitted. Let us consider the particular case in which
one has a head-on collision of two particles, particle $1$ with mass
$m_1$ and Lorentz factor $\gamma_1$, and particle $2$ with mass
$m=m_2$ with Lorentz factor $\gamma_2$, colliding to form a particle
at rest.  Without loss of generality, one may orient the axis so that
the motion is in the $(x_{D-1},x_D)$ plane, and the $x_D$ axis is the
radiation direction (see Appendix \ref{apendice}). We then have
\begin{eqnarray}
P_{1}=
\gamma_1m_1 (1,0,0,...,v_1\sin\theta_1,v_1\cos\theta_1)\,\,;\,\,\,\,\,
P'_1=(E'_{1},0,0,...,0,0)
\label{momenta1}
\\
P_{2}=
\gamma_2m_2 (1,0,0,...,-v_2\sin\theta_1,-v_2\cos\theta_1)\,\,;\,\,\,\,\,
P'_2=(E'_{2},0,0,...,0,0).
\label{momenta2}
\end{eqnarray}
Momentum conservation leads to the additional relation
$\gamma_1m_1v_1=\gamma_2m_2v_2$.  Replacing (\ref{momenta1}) and
(\ref{momenta2}) in the energy-momentum tensor
(\ref{enmomtenpointpctles}) and using (\ref{powerwavezone}) we find
\begin{equation}
\frac{d^2E}{d\omega d\Omega}=\frac{2{\cal G}}{(2\pi)^{D-2}}
\frac{D-3}{D-2}\frac{\gamma_{1}^2m_{1}^2v_{1}^2(v_1+v_2)^2
\sin{\theta_1}^4}{(1-v_1\cos\theta_1)^2(1+v_2\cos\theta_1)^2}\times \omega^{D-4}\,.
\label{energypersolidanglefreqinstcol}
\end{equation}
We see that the for arbitrary (even) $D$ the spectrum is not
flat. Flatness happens only for $D=4$. For any $D$ the total energy,
integrated over all frequencies would diverge so one needs a cutoff
frequency which shall depend on the particular problem under
consideration.  Integrating (\ref{energypersolidanglefreqinstcol})
from $\omega=0$ to the cutoff frequency $\omega_c$ we have
\begin{equation}
\frac{dE}{d\Omega}=\frac{2{\cal G}}{(2\pi)^{D-2}}
\frac{1}{D-2}\frac{\gamma_{1}^2m_{1}^2v_{1}^2(v_1+v_2)^2
\sin{\theta_1}^4}{(1-v_1\cos\theta_1)^2(1+v_2\cos\theta_1)^2}\times \omega_{c}^{D-3}\,.
\label{energypersolidangleinstcol}
\end{equation}
Two limiting cases are of interest here, namely (i) the collision
between identical particles and (ii) the collision between a light
particle and a very massive one.  In case (i) replacing $m_1=m_2=m$,
$v_1=v_2=v$, eq. (\ref{energypersolidangleinstcol}) gives
\begin{equation}
\frac{dE}{d\Omega}=\frac{8{\cal G}}{(2\pi)^{D-2}}
\frac{1}{D-2}\frac{\gamma^2m^2v^4
\sin{\theta_1}^4}{(1-v^2\cos^2\theta_1)^2}\times \omega_{c}^{D-3}\,.
\label{energypersolidangleinstcolident}
\end{equation}
In case (ii) considering $m_1\gamma_1 \equiv m\gamma <<m_2\gamma_2$, $v_1\equiv v>>v_2$,
eq. (\ref{energypersolidangleinstcol}) yields
\begin{equation}
\frac{dE}{d\Omega}=\frac{2{\cal G}}{(2\pi)^{D-2}}
\frac{1}{D-2}\frac{\gamma^2m^2v^4
\sin{\theta_1}^4}{(1-v\cos\theta_1)^2}\times \omega_{c}^{D-3}\,.
\label{energypersolidangleinstcoliheavylight}
\end{equation}
Notice that the technique just described is expected to break down if
the velocities involved are very low, since then the collision would not
be instantaneous. In fact a condition for this method to work would can be stated

 Indeed, one can see from eq.
(\ref{energypersolidangleinstcol}) that if $v\rightarrow 0$,
$\frac{dE}{d\omega}\rightarrow 0$, even though we know (see Subsection
(\ref{HeadonCollision})) that $\Delta E \neq 0$. In any case, if the
velocities are small one can use the quadrupole formula instead.

\subsection{Applications:
The Cutoff Frequency when one of the Particles is
a Black Hole and Radiation from Black Hole Pair Creation}

\subsubsection{The Cutoff Frequency when one of the Head-on Colliding 
Particles is a Black Hole}

We shall now restrict ourselves to the case (ii) of last subsection, 
in which at least one of the particles participating in the collision is a 
massive black hole, with mass $M>>m$ (where we have put $m_1=m$ and $m_2=M$). 
 Formulas
(\ref{energypersolidangleinstcol})-
(\ref{energypersolidangleinstcoliheavylight})
are useless unless one is able to determine the cutoff frequency
$\omega_c$ present in the particular problem under consideration.  In
the situation where one has a small particle colliding at high
velocities with a black hole, it has been suggested by Smarr
\cite{wein2} that the cutoff frequency should be $\omega_c \sim 1/2M$,
presumably because the characteristic collision time is dictated by
the large black hole whose radius is $2M$.  Using this cutoff he finds
\begin{equation}
\Delta E_{\rm Smarr}\sim 0.2 \gamma^2 \frac{m^2}{M}. 
\label{smarr}
\end{equation}
The exact result, using a relativistic perturbation approach which
reduces to the numerical integration of a second order differential
equation (the Zerilli wavefunction), has been given by Cardoso and
Lemos \cite{cardosolemos}, as
\begin{equation}
\Delta E_{\rm exact} = 0.26 \gamma^2 \frac{m^2}{M}. 
\label{exact}
\end{equation}
This is equivalent to saying that $\omega_c =\frac{0.613}{M} \sim \frac{1}{1.63
M}$, and so it looks like the cutoff is indeed the inverse of the
horizon radius.  However, in the numerical work by Cardoso and Lemos,
it was found that it was not the presence of an horizon that
contributed to this cutoff, but the presence of a potential barrier $V$
outside the horizon. By decomposing the field in tensorial spherical
harmonics with index $l$ standing for the angular quantum number, we
found that for each $l$, the spectrum is indeed flat (as predicted by
eq.  (\ref{energypersolidangleinstcol}) for $D=4$), until a cutoff
frequency $\omega_{c_l}$ which was numerically equal to the lowest
gravitational quasinormal frequency $\omega_{\rm QN}$. For $\omega>
\omega_{c_l}$ the spectrum decays exponentially. This behavior is
illustrated in Fig. 1.  
\begin{figure}
\centerline{\includegraphics[width=10 cm,height=6.5 cm]
{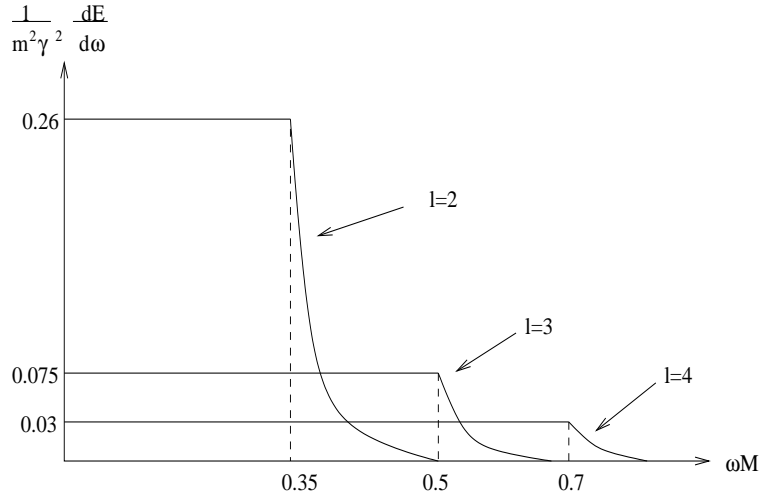}}
\caption{The energy spectra as a function of the angular number $l$,
for a highly relativistic particle falling into a $D=4$ Schwarzschild
black hole \cite{cardosolemos}. The particle begins to fall with a
Lorentz factor $\gamma$. Notice that for each $l$ there is a cutoff
frequency $\omega_{c_l}$ which is equal to the quasinormal frequency
$\omega_{\rm QN}$ after which the spectrum decays exponentially. So it
is clearly seen that $\omega_{\rm QN}$ works as a cutoff
frequency. The total energy radiated is a given by a sum over $l$,
which is the same as saying that the effective cutoff frequency is
given by a weighted average of the various $\omega_{c_l}$.}
\label{fig:1}
\end{figure}
The quasinormal frequencies \cite{kokkotas} are
those frequencies that correspond to only outgoing waves at infinity
and only ingoing waves near the horizon.  As such the gravitational
quasinormal frequencies will in general have a real and an imaginary
part, the latter denoting gravitational wave emission and therefore a
decay in the perturbation.  There have been a wealth of works dwelling
on quasinormal modes on asymptotically flat spacetimes
\cite{kokkotas}, due to its close connection with gravitational wave
emission, and also on non-asymptotically flat spacetimes, like
asymptotically anti-de Sitter \cite{horowitz} or asymptotically de
Sitter \cite{abdalla} spacetimes, mainly due to the AdS/CFT and dS/CFT
\cite{maldacena} correspondence conjecture.
We argue here that it is indeed the quasinormal frequency that
dictates the cutoff, and not the horizon radius.  For $D=4$ it so
happens that the weighted average of $\omega_{c_l}$ is
$\frac{0.613}{M}$, which, as we said, is quite similar to
$r_+=\frac{1}{2M}$.  The reason for the cutoff being dictated by the
quasinormal frequency can be understood using some WKB
intuition. The presence of a potential barrier outside the horizon
means that waves with some frequencies get reflected back on the
barrier while others can cross. Frequencies such that $\omega^2$ is
lower than the maximum barrier height $V_{\rm max}$ will be reflected
back to infinity where they will be detected. However, frequencies
$\omega^2$ larger than the maximum barrier height cross the barrier
and enter the black hole, thereby being absorbed and not contributing
to the energy detected at infinity.  So only frequencies $\omega^2$
lower than this maximum barrier height are detected at infinity.  It
has been shown \cite{schutz2} that the gravitational quasinormal
frequencies are to first order equal to the square root of the maximum
barrier height.  In view of this picture, and considering the physical
meaning of the cutoff frequency, it seems quite natural to say that
the cutoff frequency is equal to the quasinormal frequency. If the
frequencies are higher than the barrier height, they don't get
reflected back to infinity.
This discussion is very important to understand how the total energy
varies with the number $D$ of dimensions. In fact, if we set $\omega_c
\sim \frac{1}{r_+}$, we find that the total energy radiated decreases
rapidly with the dimension number, because $r_+$ increases rapidly
with the dimension. This conflicts with recent results
\cite{eardley,yoshino}, which using shock waves that describe boosted
Schwarzschild black holes, and searching for apparent horizons,
indicate an increase with $D$.
So, we need the gravitational quasinormal frequencies for higher
dimensional Schwarzschild black holes. 
To arrive at an wave equation for gravitational perturbations of
higher dimensional Schwarzschild black holes, and therefore to compute
its gravitational quasinormal frequencies, one needs to decompose
Einstein's equations in D-dimensional tensorial harmonics, which would
lead to some quite complex expressions.  It is not necessary to go
that far though, because one can get an idea of how the gravitational
quasinormal frequencies vary by searching for the quasinormal
frequencies of scalar perturbations, and scalar quasinormal
frequencies are a lot easier to find.  One hopes that the scalar
frequencies will behave with $D$ in the same manner as do the
gravitational ones.  Scalar perturbations in $D$-dimensional
Schwarzschild spacetimes obey the wave equation (consult
\cite{cardosolemos3} for details)
\begin{equation}
\frac{\partial^{2} \phi(\omega,r)}{\partial r_*^{2}} +
\left\lbrack\omega^2-V(r)\right\rbrack
\phi(\omega,r)=0 \,.
\label{scalwavequat}
\end{equation}
The potential $V(r)$ appearing in equation (\ref{scalwavequat}) is given by
\begin{equation}
V(r)=
f(r)\left\lbrack\frac{a}{r^2}+
\frac{(D-2)(D-4)f(r)}{4r^2}+\frac{(D-2)f'(r)}{2r}\right\rbrack \,,
\label{potential}
\end{equation}
where $a=l(l+D-3)$ is the eigenvalue of the Laplacian on the
hypersphere $S^{D-2}$, the tortoise coordinate $r_*$ is defined as
$\frac{\partial r}{\partial r_*}=f(r)=\left(1-\frac{16\pi {\cal G}
M}{(D-2)\Omega_{D-2}}\frac{1}{r^{D-3}}\right)$, and $f'(r)=\frac{df(r)}{dr}$.
We have found the quasinormal frequencies of spherically symmetric
($l=0$) scalar perturbations, by using a WKB approach developed by
Schutz, Will and collaborators \cite{schutz2}.  The results are
presented in Table 2, where we also show the maximum barrier height of
the potential in eq. (\ref{potential}), as well as the horizon radius.
\vskip 1mm
\begin{table}
\caption{\label{tab:zfl} The lowest scalar quasinormal frequencies for
spherically symmetric ($l=0$) scalar perturbations of higher dimensional
Schwarzschild black holes, obtained using a WKB method \cite{schutz2}.
Notice that the real part of the quasinormal frequency is always the
same order of magnitude as the square root of the maximum barrier height. 
We show also the maximum barrier height as well as the horizon radius as a
function of dimension $D$. The mass $M$ of the black hole has been set
to 1.}
\begin{ruledtabular}
\begin{tabular}{lllll}  \hline
$D$&${\rm Re}[\omega_{QN}]$:&${\rm Im}[\omega_{QN}]:$&$\sqrt{\rm V_{\rm max}}$:&$1/r_+$:\\ \hline
4   &  0.10      & -0.12         &  0.16 &  0.5 \\ \hline 
6   &  1.033     &  -0.713      &  1.441   &  1.28  \\ \hline 
8   &  1.969      &  -1.023       &  2.637      &  1.32   \\ \hline 
10  &  2.779     &   -1.158     &  3.64   &  1.25   \\ \hline 
12  &  3.49     &    -1.202     &  4.503   &  1.17 \\ \hline 
\end{tabular}
\end{ruledtabular}
\end{table}
\vskip 1mm
The first thing worth noticing is that the real part of the scalar
quasinormal frequency is to first order reasonably close to the square root
of the maximum barrier height $\sqrt{V_{\rm max}}$, supporting the
previous discussion.  Furthermore, the scalar quasinormal frequency
grows more rapidly than the inverse of the horizon radius
$\frac{1}{r_+}$ as one increases $D$. In fact, the scalar quasinormal
frequency grows with $D$ while the horizon radius $r_+$ gets smaller.
Note that from pure dimensional arguments, for fixed $D$,
$\omega \propto \frac{1}{r_+}$.
The statement here is that the constant of proportionality depends on
the dimension $D$, more explicitly it grows with $D$,  and can be found 
from Table 2.
Assuming that the gravitational quasinormal frequencies will have the
same behavior (and some very recent studies \cite{quasispectrum}
relating black hole entropy and damped quasinormal frequencies seem to
point that way), the total energy radiated will during high-energy
collisions does indeed increase with $D$, as some studies
\cite{eardley,yoshino} seem to indicate.
\subsubsection{The gravitational 
energy radiated during black hole pair creation}
As a new application of this instantaneous collision formalism, we
will now consider the gravitational energy released during the quantum
creation of pairs of black holes, a process which as far as we know
has not been analyzed in the context of gravitational wave emission,
even for $D=4$.  It is well known that vacuum quantum fluctuations
produce virtual electron-positron pairs. These pairs can become real
\cite{schwinger} if they are pulled apart by an external electric
field, in which case the energy for the pair materialization and
acceleration comes from the external electric field energy.  Likewise,
a black hole pair can be created in the presence of an external field
whenever the energy pumped into the system is enough in order to make
the pair of virtual black holes real (see Dias
\cite{diaslemos2} for a review on black hole pair creation). If one
tries to predict the spectrum of radiation coming from pair creation,
one expects of course a spectrum characteristic of accelerated masses
but one also expects that this follows some kind of signal indicating
pair creation. In other words, the process of pair creation itself,
which involves the sudden creation of particles, must imply
emission of radiation. It is this phase we shall focus on, forgetting
the subsequent emission of radiation caused by the acceleration.

Pair creation is a pure quantum-mechanical process in nature, with no
classical explanation. But given that the process does occur, one may
ask about the spectrum and intensity of the radiation accompanying
it. The sudden creation of pairs can be viewed for our purposes as an
instantaneous creation of particles (i.e., the time reverse
process of instantaneous collisions), the violent acceleration of
particles initially at rest to some final velocity in a very short
time, and the technique described at the beginning of this section
applies. This is quite similar to another pure quantum-mechanical
process, the beta decay.  
The electromagnetic radiation emitted during beta decay has been
computed classically by Chang and Falkoff \cite{chang} and is also
presented in Jackson \cite{jackson}.  The classical calculation is
similar in all aspects to the one described in this section (the
instantaneous collision formalism) assuming the sudden
acceleration to energies $E$ of a charge initially at rest, and
requires also a cutoff in the frequency, which has been assumed to be
given by the uncertainty principle $\omega_c \sim \frac{E}{\hbar}$.
Assuming this cutoff one finds that the agreement between the
classical calculation and the quantum calculation \cite{chang} is
extremely good (specially in the low frequency regime), and more
important, was verified experimentally.
Summarizing, formula (\ref{energypersolidangleinstcolident}) also describes
the gravitational energy radiated when two black holes, each with mass
$m$ and energy $E$ form through quantum pair creation.  The typical
pair creation time can be estimated by the uncertainty principle
$\tau_{\rm creation}\sim \hbar/E \sim \frac{\hbar}{m\gamma}$, and thus
we find the cutoff frequency as 
\begin{equation}
\omega_c\sim \frac{1}{\tau_{\rm creation}}\sim \frac{m\gamma}{\hbar}\,.  
\label{cutofffrequency}
\end{equation}
Here we would
like to draw the reader's attention to the fact that the units of
Planck's constant $\hbar$ change with dimension number $D$: according
to our convention of setting $c=1$ the units of $\hbar$ are
$[M]^{\frac{D-2}{D-3}}$.  With this cutoff, we find the spectrum of the
gravitational radiation emitted during pair creation to be given by
(\ref{energypersolidanglefreqinstcol}) with $m_1=m_2$ and $v_1=v_2$
(we are considering the pair creation of two identical black holes):
\begin{equation}
\frac{d^2E}{d\omega d\Omega}=\frac{8{\cal G}}{(2\pi)^{D-2}}
\frac{D-3}{D-2}\frac{\gamma^2m^2v^4
\sin{\theta_1}^4}{(1-v^2\cos^2\theta_1)^2}\times \omega^{D-4}\,,
\label{specpaircreat}
\end{equation}
and the total frequency integrated energy per solid angle is
\begin{equation}
\frac{dE}{d\Omega}=\frac{8{\cal G}}{(2\pi)^{D-2}(D-2)}\frac{v^4
\sin{\theta_1}^4}{(1-v^2\cos^2\theta_1)^2}\times \frac{(m \gamma)^{D-1}}{\hbar^{D-3}}\,.
\label{specpaircreatintfreq}
\end{equation}
For example, in four dimensions and for pairs with $v \sim 1$ one
obtains 
\begin{equation}
\frac{dE}{d\omega} =\frac{4{\cal G}}{\pi} \gamma^2m^2 \,,
\label{specpair4d}
\end{equation}
and will have for the total energy radiated during production itself,
using the cutoff frequency (\ref{cutofffrequency})
\begin{equation}
\Delta E =\frac{4{\cal G}}{\pi} 
\frac{\gamma^3m^3}{\hbar} \,.
\label{totenergpair4d}
\end{equation}
This could lead, under appropriate numbers of $m$ and $\gamma$ to huge
quantities. Although one cannot be sure as to the cutoff frequency,
and therefore the total energy (\ref{totenergpair4d}), it is extremely
likely that, as was verified experimentally in beta decay, the zero
frequency limit, eq. (\ref{specpair4d}), is exact.

\section{Summary and discussion}
We have developed the formalism to compute gravitational wave
generation in higher $D$ dimensional spacetimes, with $D$
even. Several examples have been worked out, and one cannot help the
feeling that our apparently four dimensional world is the best one to
make predictions about the intensity of gravitational waves in
concrete situations, in the sense that a small variation of parameters
leads in high $D$ to a huge variation of the energy radiated.  A lot
more work is still needed if one wants to make precise predictions
about gravitational wave generation in $D$ dimensional spacetimes.
For example, it would be important to find a way to treat
gravitational perturbations of higher dimensional Schwarzschild black
holes.  One of the examples worked out, the gravitational radiation
emitted during black hole pair creation, had not been previously
considered in the literature, and it seems to be a good candidate,
even in $D=4$, to radiate intensely through gravitational waves.

\section*{Acknowledgements}
This work was partially funded by Funda\c c\~ao para a
Ci\^encia e Tecnologia (FCT)- Portugal through project PESO/PRO/2000/4014. V.C.
and \'{O}. D. also acknowledge finantial support from FCT through PRAXIS XXI
programme.  J. P. S. L. thanks Observat\'orio Nacional do Rio de
Janeiro for hospitality.
\newpage
\appendix
\section{Spherical Coordinates in ($\bm D-1$)-dimensions}
\label{apendice}
In this appendix we list some important formulas and results used
throughout this paper.  We shall first present the transformation
mapping a ($D-1$) cartesian coordinates, ($x_1,x_2,x_3,...,x_{D-1}$)
onto ($D-1$) spherical coordinates,
($r,\theta_1,\theta_2,...,\theta_{D-2}$).
The transformation reads
\begin{eqnarray*}
x_1=r\sin\theta_1\sin\theta_2...\sin\theta_{D-2}\\
x_2=r\sin\theta_1\sin\theta_2...\sin\theta_{D-3}\cos\theta_{D-2}\\
:\\
x_i=r\sin\theta_1\sin\theta_2...\sin\theta_{D-i-1}\cos\theta_{D-i}\\
:\\
x_{D-1}=r\cos\theta_1
\label{transformation}
\end{eqnarray*}
The Jacobian of this transformation is 
\begin{equation}
J=r^{D-2}\sin\theta_1^{D-3}
\sin\theta_2^{D-4}...\sin\theta_i^{D-i-2}...\sin\theta_{D-3}\,,
\label{jacobian}
\end{equation}
and the volume element becomes
\begin{equation}
d^{D-1}{\bf x}= 
J dr d\theta_1 d\theta_2...d\theta_{D-2}=r^{D-2}dr d\Omega_{D-2}\,,
\label{volumeelement}
\end{equation}
where 
\begin{equation} 
d\Omega_{D-2}=\sin\theta_1^{D-3}\sin\theta_2^{D-4}...
\sin\theta_{D-3}d\theta_1d\theta_2...d\theta_{D-2}.
\label{solidangle}
\end{equation}
is the element of the ($D-1$) dimensional solid angle.
Finally, using \cite{grad}
\begin{equation}
\int_{0}^{\pi} \sin\theta^n=\sqrt{\pi} \frac{\Gamma[(n+1)/2]}{\Gamma[(n+2)/2]}\,,
\label{sinintegral}
\end{equation}
this yields
\begin{equation}
\Omega_{D-2}=\frac{2\pi^{(D-1)/2}}{\Gamma[(D-1)/2]}.
\label{integratedsolidangle}
\end{equation}
Here, $\Gamma[z]$ is the Gamma function, whose definition and
properties are listed in \cite{stegun}. In this work the main
properties of the Gamma function which were used are
$\Gamma[z+1]=z\Gamma[z]$ and $\Gamma[1/2]=\sqrt{\pi}$.


\end{document}